\documentclass[12pt, preprint]{revtex4-1}
\usepackage{graphicx}
\usepackage[OT1]{fontenc}
\usepackage[utf8]{inputenc}
\usepackage{bm,amsmath,amssymb,amsbsy,amstext}
\usepackage[colorlinks=false]{hyperref}
\usepackage{xcolor}

\begin{document}

\title{Consistent simulation of capacitive radio-frequency discharges and external matching networks}
\author{Frederik Schmidt}
\affiliation{%
Institute of Theoretical Electrical Engineering, Ruhr University Bochum, Bochum, 44780, Germany
}%

\author{Thomas Mussenbrock}
\affiliation{%
Electrodynamics and Physical Electronics Group, Brandenburg University of Technology Cottbus-Senftenberg, Cottbus, 03046, Germany
}%
\author{Jan Trieschmann}
\affiliation{%
Electrodynamics and Physical Electronics Group, Brandenburg University of Technology Cottbus-Senftenberg, Cottbus, 03046, Germany
}%
\date{\today}

\begin{abstract}

\medskip

External matching networks are crucial and necessary for operating capacitively coupled plasmas in order to maximize the absorbed power. Experiments show that external circuits in general heavily interact with the plasma in a nonlinear way. This interaction has to be taken into account in order to be able to design suitable networks, e.g., for plasma processing systems. For a complete understanding of the underlying physics of this coupling, a nonlinear simulation approach which considers both the plasma and the circuit dynamics can provide useful insights. In this work, the coupling of an equivalent circuit plasma model and an electric external circuit composed of lumped elements is discussed. The plasma model itself is self-consistent in the sense that the plasma density and the electron temperature is calculated from the absorbed power based on a global plasma chemistry model. The approach encompasses all elements present in real plasma systems, i.e., the discharge itself, the matching network, the power generator as well as stray loss elements. While the main results of this work is the conceptual approach itself, at the example of a single-frequency capacitively coupled discharge its applicability is demonstrated. It is shown that it provides an effective and efficient way to analyze and understand the nonlinear dynamics of real plasma systems and, furthermore, may be applied to synthesize optimal matching networks.
\end{abstract}

\maketitle

\section{\label{sec:introduction} Introduction}

Capacitively coupled plasmas (CCPs) operated at radio frequencies (RFs) are powered by generators that are connected to the driven electrode via an external electrical   circuit.\cite{lieberman_principles_2005,chabert_physics_2011} These circuits depend on the specific setup and may among other elements include power lines, matching networks, frequency filters. Since the plasma is typically a nonlinear load, its interaction with the external circuit is not easily predictable. Especially at low pressures $p<10$~Pa, harmonics in the plasma current at the plasma series resonance can have a strong influence on the plasma characteristics.\cite{mussenbrock_nonlinear_2007,mussenbrock_enhancement_2008, czarnetzki_self-excitation_2006, lieberman_effects_2008, miller_electrical_1992, ziegler_temporal_2009, yamazawa_effect_2009} It has been observed by several researchers in the field that these harmonics interact heavily with the external circuit connected to the driven discharge electrode,\cite{miller_electrical_1992, yamazawa_effect_2009, rauf_effect_1998, sobolewski_electrical_1999, kaiser_optimization_2013}, e.g., reacting to simple changes such as a varied cable length.\cite{miller_electrical_1992} Yamazawa et al. demonstrated that this interaction can be used to influence the electron density and its homogeneity by changing the composition of the harmonics with an LC unit parallel to the plasma.\cite{yamazawa_control_2007, yamazawa_electrode_2015}

Since analytic predictions of these almost always nonlinear effects are hard to obtain, an effective numerical simulation that takes both plasma and external circuit into account may provide important insights (e.g., to design suitable matching and filter networks). Certain assumptions about the plasma states are, however, often necessary. Verboncoeur et al. proposed a method for coupling external circuits to a particle in cell (PIC) simulation via conservation of charge.\cite{verboncoeur_simultaneous_1993} The external circuit is incorporated by setting up the differential equations following Kirchhoff's circuit laws and solving them numerically simultaneously with the PIC simulation. Rauf and Kushner analyzed the plasma circuit interaction by coupling a lumped circuit to a plasma simulated with HPEM.\cite{rauf_effect_1998} They observed that the specific design of a matching network has a significant influence on the plasma and, therefore, the whole system of interest. 

In this work, we propose a model of a real plasma system and a simulation method in which the plasma itself is modeled by a nonlinear equivalent circuit consistently coupled to a global chemistry model. Therefore, the problem reduces to a circuit simulation for which several solution tools are available. These tools, one of which is SPICE, solve the system of differential equations using a netlist of the circuit.\cite{nagel_spice_1973} Here, we make use of the open-source software ngSPICE.\cite{vogt_ngspice_2017} The differential equations, hence, do not need to be set up and solved ``by hand'', which is especially cumbersome for large circuits. As aforementioned, the proposed simulation algorithm self-consistently calculates the electron temperature and the plasma density based on the absorbed power and parameters of the discharge. The functionality of the proposed method is demonstrated by simulating a real generator with an impedance matching network attached to the plasma model (including losses in the network and the reactor chamber). 

\section{\label{sec:method} Plasma Model and Simulation Algorithm}

In order to describe efficiently the nonlinear dynamics of real world RF plasma systems on the high-frequency timescale and to understand the interaction between the plasma and the external network, a lumped element circuit model for the plasma is established. 
In this work, the model is based on considerations introduced and discussed in~\cite{mussenbrock_enhancement_2008,mussenbrock_nonlinear_2006,lieberman_effects_2008,ziegler_temporal_2009}. It additionally allows to consistently calculate the electron temperature and density from particle and energy conservation in the system.\cite{lieberman_principles_2005}

The plasma is divided into the plasma bulk and the two sheaths: One at the driven electrode and one at the grounded electrode and reactor chamber wall. A generalized Ohm's law based on the momentum equation for electrons of the form
\begin{align}
	\frac{\partial \vec{j}}{\partial t}=\frac{e^2 n}{m_\mathrm{e}} \vec{E} - \nu_\mathrm{eff} \vec{j} \label{eq:ohms_law}
\end{align} 
is used to model the bulk on the RF timescale. $\vec{j}$ is the current density, $\vec{E}$ the electric field, $n$  the plasma density, and $m_\mathrm{e}$ the electron mass. The effective collision frequency $\nu_\mathrm{eff} = \nu_\mathrm{m} + \bar{v}_\mathrm{e}/l_\mathrm{B}$ accounts for both ohmic heating by incorporating the momentum transfer collision frequency $\nu_\mathrm{m}$ and stochastic heating in form of the second term. Therein, $\bar{v}_\mathrm{e} = \left( 8 k_\mathrm{B} T_\mathrm{e}/\pi m_\mathrm{e} \right)^\frac{1}{2}$ is the mean thermal speed, $T_\mathrm{e}$ is the electron temperature, $l_\mathrm{B}$ is the bulk length, and $k_\mathrm{B}$ is the Boltzmann constant. Assuming a homogeneous cylindrical discharge, integration over the electrode area $A_\mathrm{E}$ and the bulk length $l_\mathrm{B}$ respectively leads to a scalar version of equation~\eqref{eq:ohms_law}. This can be written in the form $V_\mathrm{Bulk} =  R_\mathrm{pl} I_\mathrm{pl} + L_\mathrm{pl}  \partial_t I_\mathrm{pl}$, with $V_\mathrm{Bulk}$ the voltage dropping over the bulk and $I_\mathrm{pl}$ the current flowing through the bulk and, therefore, through the whole discharge. Recapitulatory, the bulk is modeled by an inductor $L_\mathrm{pl}=l_\mathrm{B} m_\mathrm{e}/e^2 n A_\mathrm{E}$ and a resistor $R_\mathrm{pl} = \nu_\mathrm{eff} L_\mathrm{pl}$. 

\begin{figure}[t!]
	\centering
	\resizebox{8cm}{!}{		
		\includegraphics[width=8cm]{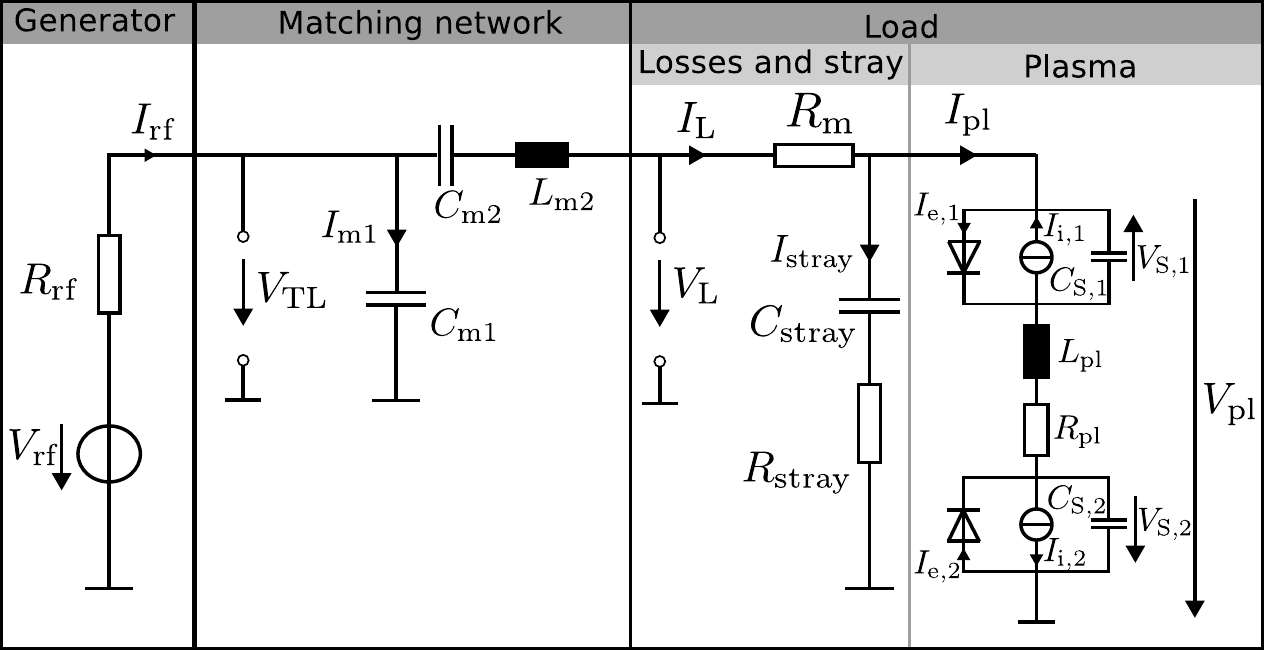}
	}
	\caption{Network that is simulated. On the right hand side is the equivalent circuit of the plasma.}
	\label{fig:setup}
\end{figure}

The sheaths of the plasma are modeled by a capacitive diode consisting of three parallel elements. Due to their high inertia, the ion flux onto the surface is assumed constant and is accordingly modeled as a constant ion current source $I_\mathrm{i,1} = A_\mathrm{E} e n u_\mathrm{B}$, with the Bohm velocity $u_\mathrm{B} = \sqrt{k_\mathrm{B} T_e/m_\mathrm{i}}$ and the ion mass $m_\mathrm{i}$. The electrons have a much lower mass $m_\mathrm{e}$ and thus depend on the time-varying sheath voltage $V_\mathrm{S}$. The electron current is then $I_\mathrm{e,1} = A_\mathrm{E} e n \bar{v}_\mathrm{e} \mathrm{exp}\left( -e V_\mathrm{S,1}/k_\mathrm{B} T_\mathrm{e} \right)$, assuming a Maxwellian electron energy distribution. Lastly, the sheath capacitance is modeled as a nonlinear capacitor $C_\mathrm{S,1}= \left( 2 e n \epsilon_0 A_\mathrm{E}^2/V_\mathrm{S,1} \right)^{\frac{1}{2}}$. The nonlinearity itself depends on the applied sheath model. In this work a matrix sheath model~\cite{lieberman_principles_2005} is used. $I_\mathrm{i,1}$, $I_\mathrm{e,1}$ and $C_\mathrm{S,1}$ are the values describing the sheath in front of the driven electrode. The values for the grounded electrode sheath are obtained by exchanging $A_\mathrm{E}$ with $A_\mathrm{G}$, consequently, $I_\mathrm{i,2} = A_\mathrm{G} e n u_\mathrm{B}$, $I_\mathrm{e,2} = A_\mathrm{G} e n \bar{v}_\mathrm{e} \mathrm{exp}\left( -e V_\mathrm{S,2}/k_\mathrm{B} T_\mathrm{e} \right)$ and $C_\mathrm{S,2}= \left( 2 e n \epsilon_0 A_\mathrm{G}^2/V_\mathrm{S,2} \right)^{\frac{1}{2}}$.
The system of differential equations for the plasma model to be solved then amounts to
\begin{eqnarray}\label{eq:diff1}
 \frac{d V_\mathrm{S,1}}{d t} &=& -C_\mathrm{S,1}^{-1}(I_\mathrm{pl} + I_\mathrm{i,1} - I_\mathrm{e,1}) \\ \label{eq:diff2} 
 \frac{d V_\mathrm{S,2}}{d t} &=& -C_\mathrm{S,2}^{-1}(-I_\mathrm{pl} + I_\mathrm{i,2} - I_\mathrm{e,2}) \\ \label{eq:diff3}
  \frac{d I_\mathrm{pl}}{d t} &=& L_\mathrm{p}^{-1}(V_\mathrm{pl} + V_\mathrm{S,1} - V_\mathrm{S,2}) - \nu_\mathrm{eff} I_\mathrm{pl},
\end{eqnarray}
with $V_\mathrm{pl}$ being the voltage dropping over the whole discharge.
This system of equations can be described by an equivalent circuit model, which is depicted on the very right hand side of figure~\ref{fig:setup}.

Since this model consists only of circuit elements, arbitrary networks using sources, resistors, capacitors, inductors and so on can be attached to the electrode and, following Kirchhoff's laws, the resulting differential equations can be solved. However, since the plasma density $n$ depends on the absorbed power $P_\mathrm{abs}$, which in return depends on the attached electric circuit, the model needs to be extended to catch this dependency. Following the argument of Lieberman~\cite{lieberman_principles_2005}, the energy lost per electron-ion pair created depends on the energy lost due to collisions and due to particles leaving the system. For the former, an equation $\mathcal{E}_\mathrm{c} = \left( K_\mathrm{iz} \mathcal{E}_\mathrm{iz} + K_\mathrm{ex} \mathcal{E}_\mathrm{ex}+ K_\mathrm{el} \mathcal{E}_\mathrm{el} \right)/K_\mathrm{iz}$  can be derived. The energy lost due to collisions $\mathcal{E}_\mathrm{c}$ depends on the rate constants $K_\mathrm{iz}, K_\mathrm{ex}$, and $K_\mathrm{el}$ for ionization, excitation and elastic collisions as well as the respective energies $\mathcal{E}_\mathrm{iz}, \mathcal{E}_\mathrm{ex}$ and $\mathcal{E}_\mathrm{el}$. The values of the rate constants depend on the chosen background gas and on the electron temperature. In this work, approximation functions provided by Gudmundsson for an argon discharge are used.\cite{gudmundsson_notitle_2002} The energy lost by particles leaving the system needs to be accounted for, for each charged species. For electrons following a Maxwellian distribution this energy can be assumed to be  $\mathcal{E}_\mathrm{e} = 2 k_\mathrm{B} T_\mathrm{e}$. Ions have an initial energy of $k_\mathrm{B} T_\mathrm{e}/2$ at the Bohm point and are then accelerated following the mean sheath voltage $\overline{V}_\mathrm{S}$ resulting in $\mathcal{E}_\mathrm{i} =  e\,\overline{V}_\mathrm{S} + k_\mathrm{B} T_\mathrm{e}/2$. Accounting for the different sheaths at the driven and grounded electrode and the electrode sizes respectively, the total ion energy lost results in $\mathcal{E}_\mathrm{i} = f_\mathrm{E} \,e\,\overline{V}_\mathrm{S,1} + f_\mathrm{G} \,e\,\overline{V}_\mathrm{S,2} + k_\mathrm{B} T_\mathrm{e}/2$ with the weighting factors $f_\mathrm{E} = A_\mathrm{E}/(A_\mathrm{E}+A_\mathrm{G})$ and $f_\mathrm{G} = A_\mathrm{G}/(A_\mathrm{E}+A_\mathrm{G})$. The absorbed power can then be calculated taking all created particles in the plasma volume $V_\mathrm{p} = A_\mathrm{E} l_\mathrm{B}$ into account, which results in
\begin{align}
	P_\mathrm{abs} = n V_\mathrm{p} n_\mathrm{g} K_\mathrm{iz} \left(  \mathcal{E}_\mathrm{c}+\mathcal{E}_\mathrm{e}+f_\mathrm{E} \mathcal{E}_\mathrm{i,E} + f_\mathrm{G} \mathcal{E}_\mathrm{i,G}\right), \label{eq:power_balance}
\end{align} 
with the neutral gas density $n_\mathrm{g}$.
Since the total number of electrons and ions created needs to be equal to the number of particles leaving the system, a condition for particle conservation
\begin{align}
	V_\mathrm{p} n_\mathrm{g} K_\mathrm{iz} = u_\mathrm{B} A 
    \label{eq:bohm_condition}
\end{align} 
can be established with the total area $A=A_\mathrm{E}+A_\mathrm{G}$. $A$ and $V_\mathrm{p}$ are geometrical values that depend on the actual setup of the plasma. Again, assuming a cylindrical discharge, both values can be calculated once the electrode area $A_\mathrm{E}$ is known. It is in this case important to keep in mind that $A_\mathrm{G}$ is not necessarily the grounded area of the reactor, but rather the sheath size in front of the grounded reactor parts and, therefore, typically smaller. Once geometrical assumptions about the discharge are made, equation \eqref{eq:bohm_condition} can be used to calculate the electron temperature with both $u_\mathrm{B}$ and $K_\mathrm{iz}$ being functions of $T_\mathrm{e}$. It is also possible to do it the other way around, i.e., to calculate $A$ and $V_\mathrm{p}$ from the electron temperature, which might be known for a specific discharge from measurements.

\begin{figure}[b!]
	\centering
	\resizebox{8cm}{!}{		
		\includegraphics[width=8cm]{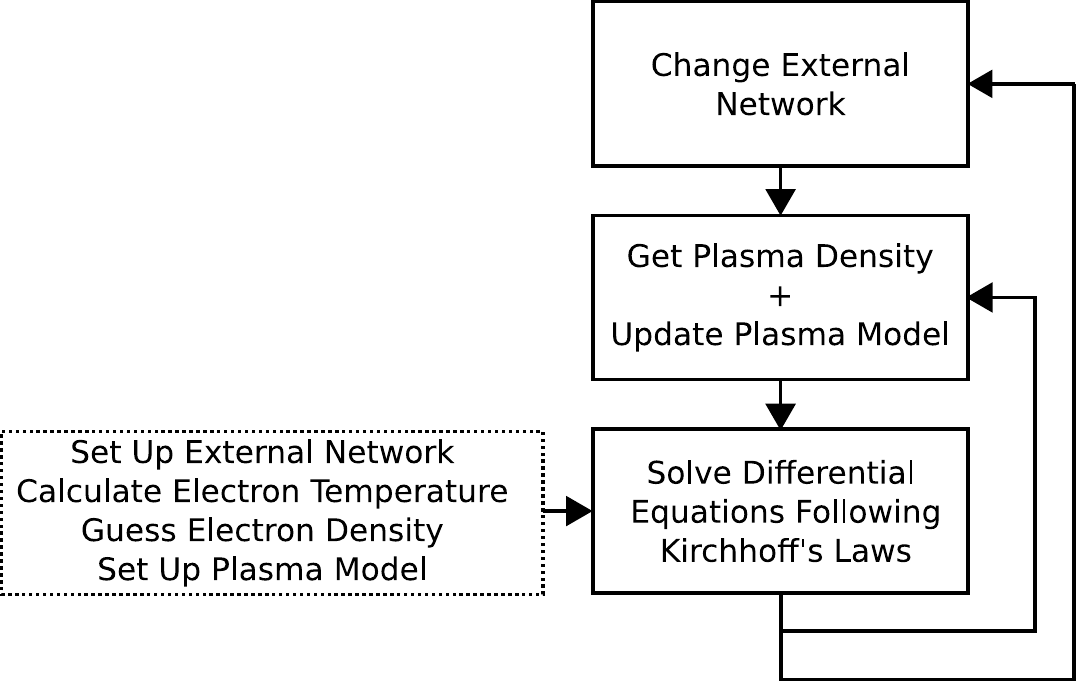}
	}
	\caption{Flowchart of the algorithm of the simulation.}
	\label{fig:algorithm}
\end{figure}

In the actual simulation algorithm depicted in figure~\ref{fig:algorithm}, equation \eqref{eq:bohm_condition} is evaluated once at the beginning of the simulation in order to set $T_\mathrm{e}$. It is assumed that the electron temperature and the geometrical parameters do not vary with the absorbed power, i.e., the plasma does not vary in size. In order to perform a transient simulation of the whole system of interest, all elements of the circuit have to be assigned specific values. In other words, the external lumped element circuit needs to be set up (e.g., generator, matching network, loss elements) and the elements of the plasma model need to be calculated. All values of the plasma elements depend on $T_\mathrm{e}$ -- which is already known at this point -- and the plasma density for which a (reasonable) value is guessed. This leads to a network with completely determined elements. Especially for large networks it is highly advisable to not set up the differential equations for the whole network by hand, but to rather make use of a circuit simulation software, for which several tools available. Due to its flexibility, being open-source and providing shared libraries, we make use of the software ngSPICE, which is based on SPICE.\cite{vogt_ngspice_2017,nagel_spice_1973} 

A transient simulation of the circuit determines all currents and voltages at each branch and node. Therefore, the absorbed power of the plasma can be calculated from the voltage $V_\mathrm{pl}$ dropping over the plasma and the current $I_\mathrm{pl}$ flowing through it in the form $P_\mathrm{abs}=\int I_\mathrm{pl} V_\mathrm{pl} dt$ (cf. figure \ref{fig:setup}). The averaged voltages across the sheaths are obtained from the simulation as well. Equation~\eqref{eq:power_balance} is then used to calculate the plasma density from these values. Afterwards, using this new value of $n$, the elements of the plasma model can be updated and the simulation can be performed again. This step is repeated until a steady state of $n$ and $P_\mathrm{abs}$ is reached. At this point, for a specific circuit, consistent values of $n$, $P_\mathrm{abs}$, $I_\mathrm{pl}$ and any other current, voltage or power of interest are obtained. Changes in the external circuit can be applied, which is the outer iteration loop of the algorithm (cf. figure \ref{fig:algorithm}), and the correct values of $n$ and $P_\mathrm{abs}$ obtained again. In this work, changes to the network are employed in the impedance matching network until the load is matched to the generator.

\section{\label{sec:results} Results and Discussion}

A typical network commonly attached to an RF plasma discharge consists of a generator (to provide power) and a matching network (to insure that a maximum amount of this power actually reaches the plasma). Such a setup is simulated in this work as depicted in figure~\ref{fig:setup}. The generator consists of a voltage source $V_\mathrm{rf} = V_0 \mathrm{cos} (\omega_1 t)$, with $V_0 = 100$~V, $\omega_1 = 2 \pi \times 13.56$~MHz, and an internal resistor $R_\mathrm{rf} = 50~\Omega$. The matching network is a typical L-type network with the tunable capacitors $C_\mathrm{m1}$ and $C_\mathrm{m2}$ and a fixed inductor $L_\mathrm{m2} = 1500$~nH. Accumulated losses of the matching network are included in the form of a resistor $R_\mathrm{m} = 0.5~\Omega$. Stray effects that occur at the reactor chamber and supply lines are modeled in the form of a resistor $R_\mathrm{stray} = 0.5~\Omega$ and a capacitor $C_\mathrm{stray} = 200$~pF. All values of the loss elements are rough estimates and intrinsically depend on the setup of the chamber, the network and supply lines, and the specific choice of matching elements. To simulate a specific setup, the losses and stray effects of the network can be measured~\cite{lieberman_principles_2005,godyak_situ_1990} and thus included in the simulation. Other sources of stray effects (e.g., in the matching network) are ignored in this work.

The discharge considered in the simulation has a driven electrode area $A_\mathrm{E} = 100~\mathrm{cm}^{2}$, a grounded area $A_\mathrm{G} = 300~\mathrm{cm}^{2}$ and a bulk length $l_\mathrm{B} = 5.7$~cm. The pressure is $p=0.66$~Pa, resulting in a plasma that is expected to be nonlinear and, therefore, to generate harmonics in the plasma current. The background gas is argon with a temperature $T_\mathrm{g} = 300$~K, which results in the neutral gas density of $n_\mathrm{g} = p/k_\mathrm{B} T_\mathrm{g} = 1.59 \times 10^{20}~\mathrm{m}^{-3}$. From these values and equation \eqref{eq:bohm_condition} an electron temperature of $T_\mathrm{e} = 4.75$~eV is calculated.

In the outer loop of the algorithm (accounting for changes in the external circuit), $C_\mathrm{m1}$ and $C_\mathrm{m2}$ are optimized in order to maximize the absorbed power of the load. This is done by performing a fast Fourier transform (FFT) of the transient load voltage $V_\mathrm{L}$ and current $I_\mathrm{L}$, and calculating the complex load impedance  $Z_\mathrm{L}(\omega_k) = \mathcal{F}\{V_\mathrm{L}\}/\mathcal{F}\{I_\mathrm{L}\}$]. At the excitation frequency $\omega_1$ with $k=1$, the matching conditions are $\mathrm{Re}\{R_\mathrm{rf} || C_\mathrm{m1}\} = \mathrm{Re}\{ Z_\mathrm{L} \}$ and $\mathrm{Im}\{ Z_\mathrm{L} \} + \mathrm{Im}\{ R_\mathrm{rf} || C_\mathrm{m1} \} = j \omega_1 L_\mathrm{m2} + 1/j \omega_1 C_\mathrm{m2}$ \cite{lieberman_principles_2005}. $C_\mathrm{m1}$ and $C_\mathrm{m2}$ can be chosen to satisfy these equations. The simulation is repeated afterwards with these new values following the algorithm depicted in figure~\ref{fig:algorithm}. Since changes in the matching network have a significant influence on the absorbed power in the load and, therefore, the plasma (which changes its elements and, accordingly, the impedance $Z_\mathrm{L}$), this step has to be repeated several times, until $C_\mathrm{m1}$ and $C_\mathrm{m2}$ reach a steady state. Typically, up to five iterations of changes in the matching network are required.

A converged and matched steady state simulation is obtained for the values $n = 1.25 \times 10^{15}~\mathrm{m}^{-3}$, $C_\mathrm{m2} = 175$~pF and $C_\mathrm{m1} = 1550$~pF. The total load at $\omega_1$, including the matching network, amounts to an impedance $Z_\mathrm{TL} = V_\mathrm{TL}/I_\mathrm{rf} = (50.01 - j 0.02)~\Omega$, which is very close to $R_\mathrm{rf} = 50~\Omega$ and, therefore, close to perfect matching to the generator. This value can be improved by using more iterations in the algorithm and especially by choosing smaller timesteps and longer simulation periods in the transient simulation of the circuit performed by ngSPICE, which would increase the numerical accuracy.

\begin{figure}[t!]
	\centering
	\resizebox{8cm}{!}{		
		\includegraphics[width=8cm]{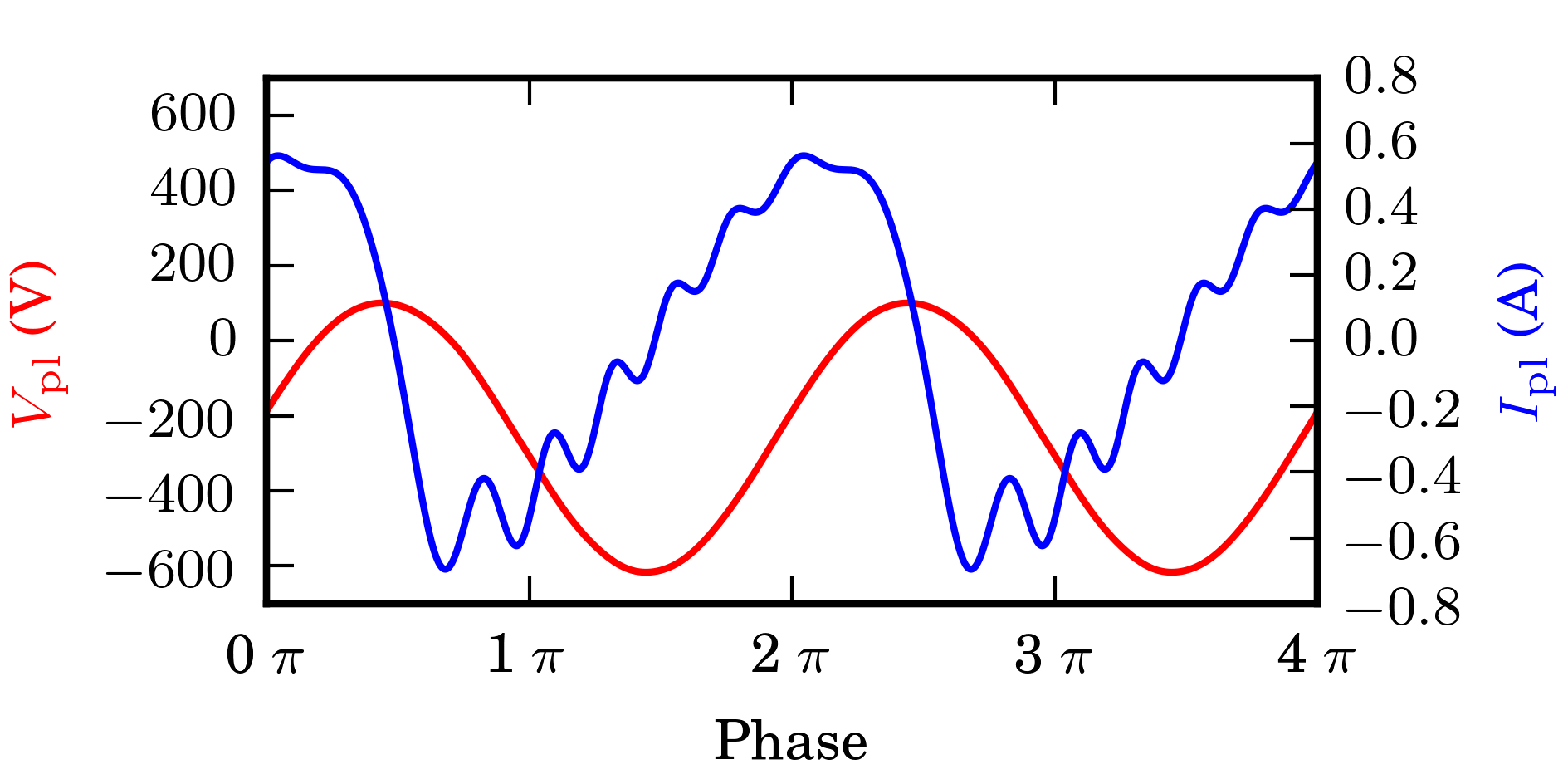}
		
	}
	\caption{Current $I_\mathrm{pl}$ flowing through the plasma and voltage $V_\mathrm{pl}$ at the driven electrode.}
	\label{fig:results_current_voltage}
\end{figure}

The resulting transient values of the current $I_\mathrm{pl}$ and the voltage $V_\mathrm{pl}$ are plotted in figure \ref{fig:results_current_voltage}. The voltage has an amplitude of about 360~V and an offset of -250~V, while being almost completely sinusoidal. The current on the other hand is composed of a variety of harmonics due to the nonlinearity of the plasma. It is particularly interesting to analyze the different harmonics at the various network nodes.

\begin{figure}[t!]
	\centering
	\resizebox{8cm}{!}{		
		\includegraphics[width=8cm]{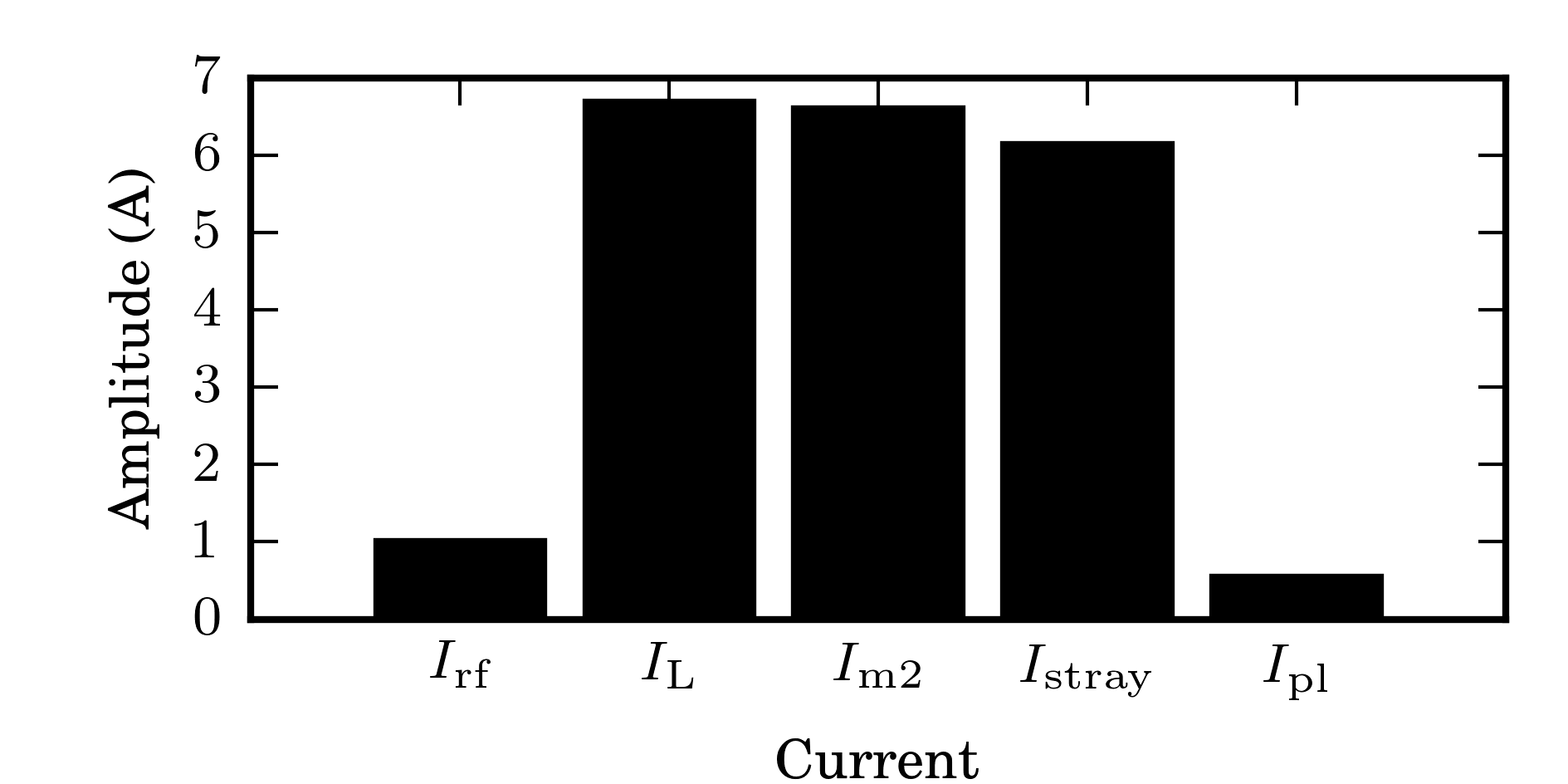}
		
	}
	\caption{Amplitudes of the currents in different branches of the network at the excitation frequency $\omega_1$.}
	\label{fig:results_current_omega_1}
\end{figure}

\begin{figure}[t!]
	\centering
	\resizebox{8cm}{!}{		
		\includegraphics[width=8cm]{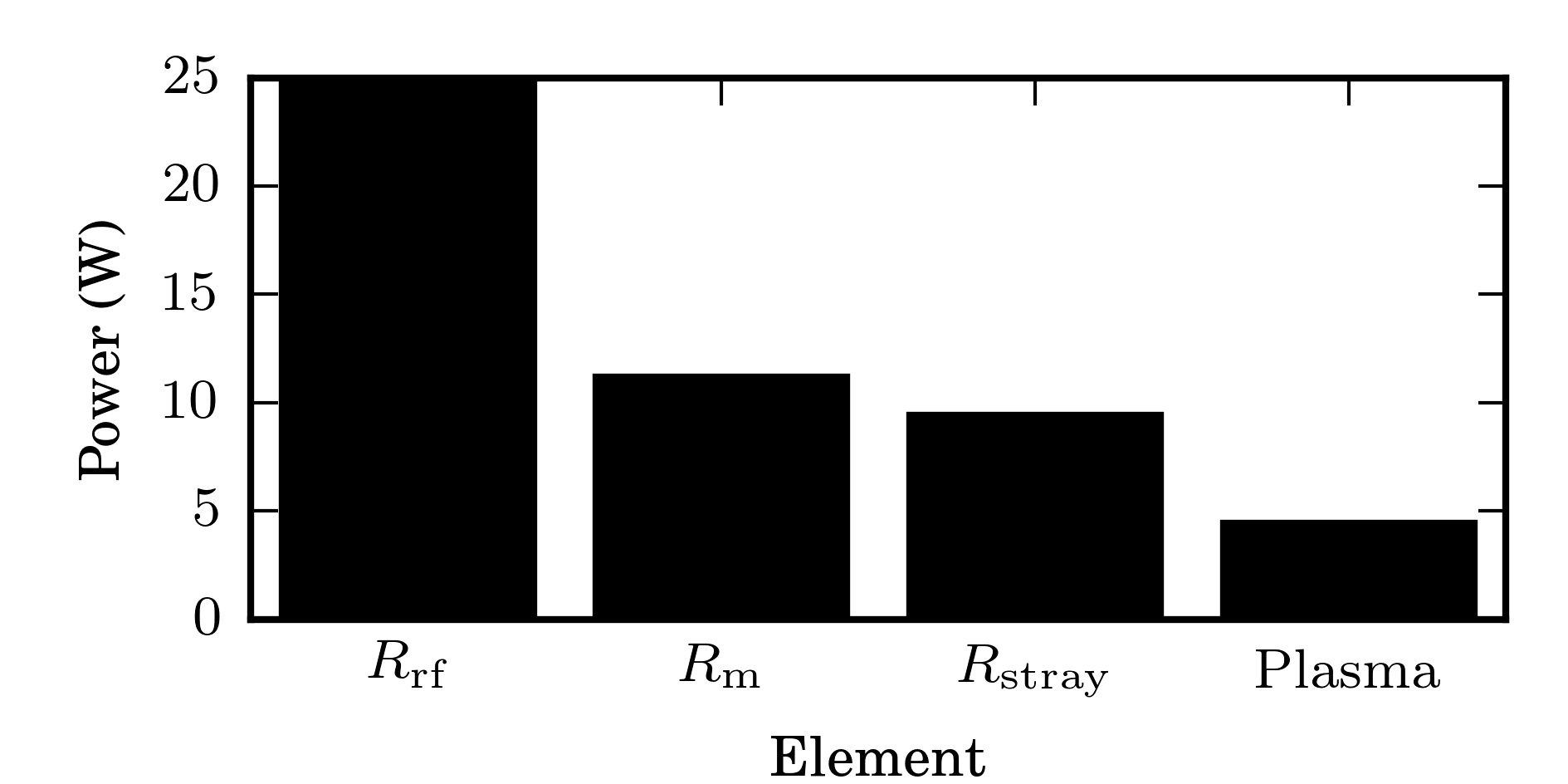}
		
	}
	\caption{Powers dissipated in the elements $R_\mathrm{rf}$, $R_\mathrm{m}$, $R_\mathrm{stray}$ and in the plasma model.}
	\label{fig:results_power_omega_1}
\end{figure}

In figure~\ref{fig:results_current_omega_1}, the amplitudes of the different currents in the system at the excitation frequency $\omega_1$ obtained by a FFT are depicted. As expected, $I_\mathrm{rf} = 1$~A, which is the same result as $I_\mathrm{rf} = |V_\mathrm{rf}/(R_\mathrm{rf} + Z_\mathrm{TL}) |$ yields for a perfectly matched load. 
The current flowing in the plasma is with $I_\mathrm{pl} \approx 0.6$~A much smaller than the currents $I_\mathrm{L} \approx 6.7$~A , $I_\mathrm{m2} \approx 6.6$~A and $I_\mathrm{stray} \approx 6.1$~A in the other branches. These currents lead to powers absorbed at different locations in the system: In the plasma, in the matching network at $R_\mathrm{m}$, in the stray line at $R_\mathrm{stray}$ and in the internal resistance $R_\mathrm{rf}$. The values are plotted in figure~\ref{fig:results_power_omega_1}. At the internal resistance $R_\mathrm{rf}$, 25~W are dissipated. Due to the load being perfectly matched, the same power is distributed over the different load elements: At the resistance $R_\mathrm{m}$ with $\sim$ 11~W, at $R_\mathrm{stray}$ with $\sim$ 9~W and at the plasma with $\sim$ 5~W. In conclusion, in this specific setup about a fifth of the power transferred to the load is dissipated in the plasma, while the rest is lost. This is mainly due to the high circular currents that flow through the matching and stray branches. In practice, the lost power leads to heating, especially in the matching network.

\begin{figure}[t!]
	\centering
	\resizebox{8cm}{!}{		
		\includegraphics[width=8cm]{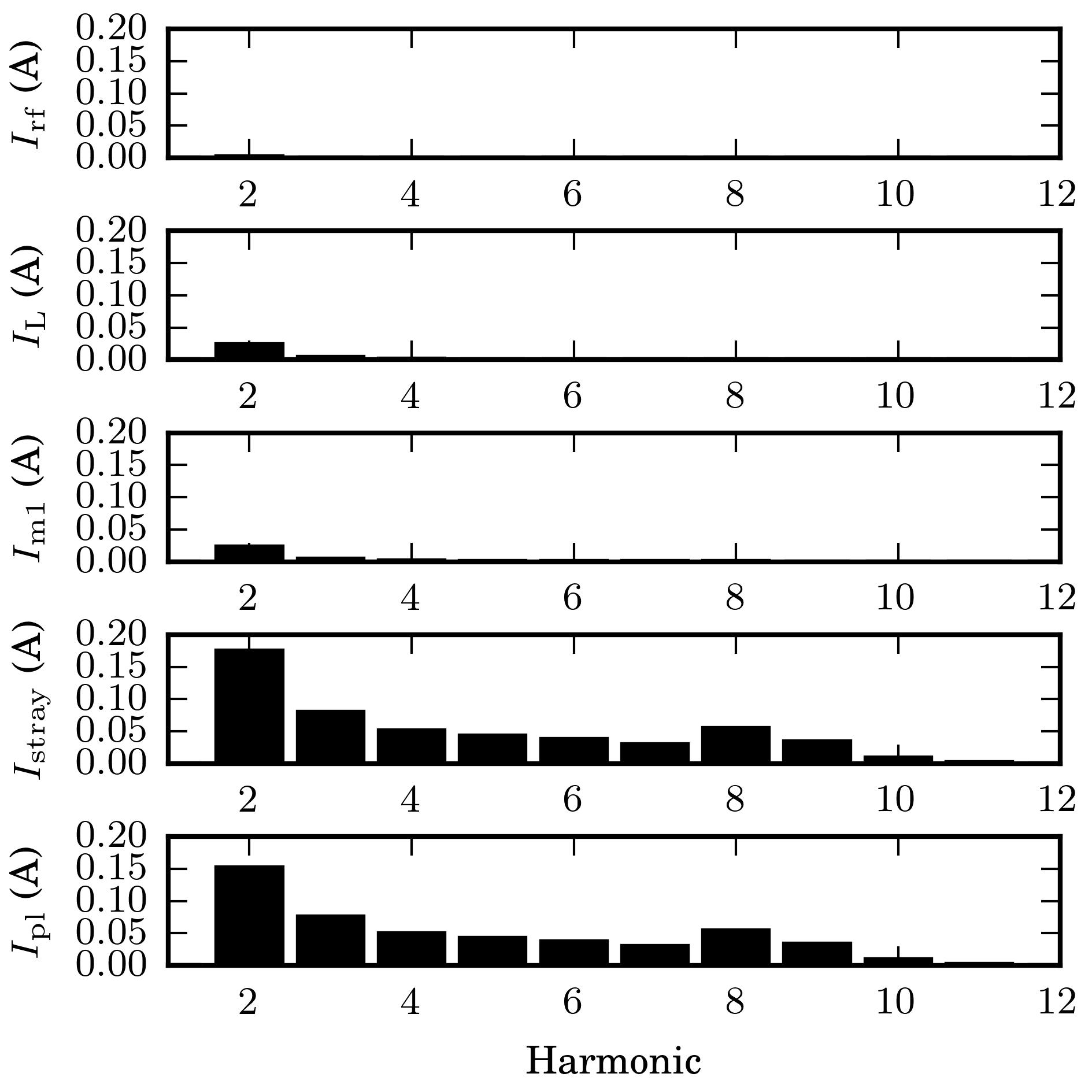}
	}
	\caption{Amplitudes of different currents in the network for different harmonics of the excitation frequency. The fundamental frequency is not included in this plot.}
	\label{fig:results_current_harmonices}
\end{figure}

The nonlinearly created harmonics in the current for which the plasma acts as a generator are another peculiarity. Leaving out the dominant current component at the fundamental excitation frequency, figure~\ref{fig:results_current_harmonices} presents the amplitudes of the different harmonics of the currents through the various branches of the system. The plasma generates a total current at $2 \omega_1$ of about 0.15~A amplitude. There exist harmonics up to the 12th harmonic with comparably small amplitudes of around 0.05~A. The stray current $I_\mathrm{stray}$ is of roughly same amplitude, while the values for $I_\mathrm{L}$ and $I_\mathrm{m1}$ are much smaller. Only the second harmonic with $\sim$ 0.02~A is clearly distinguishable in the plot. In the generator current $I_\mathrm{rf}$ even the second harmonic is almost completely suppressed. 
This effect can be understood by considering the position of the plasma in the network as a generator. In this case the three branches with $C_\mathrm{stray}$, $L_\mathrm{m2}$ and $C_\mathrm{m1}$ act as a (lossy) low-pass filter of third order with each branch suppressing high frequencies. This suppresses higher harmonics up to the point that non factually reach the ``load'' (the RF source branch). This behavior is generally desirable since higher harmonics might otherwise create problems in the form of reactive power in the generator. 

Stray effects occur in practice not in the form of a specified branch as in the simulation, but distributed over the system, such as at the coaxial cable feed and at the driven electrode~\cite{lieberman_principles_2005}. The further away from the electrode in the direction of the matching network a measurement of the current is performed, the less of the plasma-generated harmonics this current is expected to contain. Resultantly, a measurement of the plasma current that is desired to include these harmonics needs to be performed as close to the plasma as possible. 

\section{\label{sec:conclusion} Conclusion}

A simulation method for self-consistently simulating capacitively coupled plasmas with external lumped element circuits attached is developed based on an equivalent global plasma model. The model is extended to consistently include the calculation of the electron temperature and the plasma density. This is achieved based on both a particle balance and a power balance equation of the system. Using the circuit simulation software ngSPICE, a comprising simulation tool is obtained that allows for simulating a system without the need of setting up and solving the differential equations by hand. This tool is readily applicable for a study of the nonlinear circuit/plasma interaction and may also guide the design of external circuits (e.g., a matching network), because parameters like the absorbed power in the plasma, the load impedance and all the intrinsic currents and voltages in the system can easily be obtained. 

The versatility of this method is demonstrated by the simulation of a real generator, an L-type matching network and loss channels attached to the plasma. It is shown that circular currents in the matching branches arise when the matching conditions are met that lead to power dissipations in the loss elements. The nonlinearly created harmonics mostly flow through the stray reactor line and only a small amount reaches the matching network and even less the generator branch.

\section*{Acknowledgement}

This work is supported by the German Research Foundation (DFG) in the frame of the Collaborative Research Centre TRR\,87 (SFB-TR\,87). The authors thank T. Gergs from the Institute of Theoretical Electrical Engineering, Ruhr University Bochum for fruitful discussions.

\bibliographystyle{aip}

\bibliography{CCP_matching.bib}

\begin{thebibliography}{10}

\bibitem{lieberman_principles_2005}
M.~A. Lieberman and A.~J. Lichtenberg,
\newblock {\em Principles of plasma discharges and materials processing},
\newblock Wiley-Interscience, Hoboken, N.J, 2nd ed edition, 2005.

\bibitem{chabert_physics_2011}
P.~Chabert and N.~Braithwaite,
\newblock {\em Physics of radio-frequency plasmas},
\newblock Cambridge University Press, Cambridge, 2011.

\bibitem{mussenbrock_nonlinear_2007}
T.~Mussenbrock and R.~P. Brinkmann,
\newblock Plasma Sources Science and Technology {\bf 16}, 377 (2007).

\bibitem{mussenbrock_enhancement_2008}
T.~Mussenbrock, R.~P. Brinkmann, M.~A. Lieberman, A.~J. Lichtenberg, and
  E.~Kawamura,
\newblock Physical Review Letters {\bf 101}, 085004 (2008).

\bibitem{czarnetzki_self-excitation_2006}
U.~Czarnetzki, T.~Mussenbrock, and R.~P. Brinkmann,
\newblock Physics of Plasmas {\bf 13}, 123503 (2006).

\bibitem{lieberman_effects_2008}
M.~A. Lieberman, A.~J. Lichtenberg, E.~Kawamura, T.~Mussenbrock, and R.~P.
  Brinkmann,
\newblock Physics of Plasmas {\bf 15}, 063505 (2008).

\bibitem{miller_electrical_1992}
P.~A. Miller,
\newblock Electrical characterization of {RF} plasmas,
\newblock in {\em Process {Module} {Metrology}, {Control} and {Clustering}},
  pages 179--188, International Society for Optics and Photonics, 1992.

\bibitem{ziegler_temporal_2009}
D.~Ziegler, T.~Mussenbrock, and R.~P. Brinkmann,
\newblock Physics of Plasmas {\bf 16}, 023503 (2009).

\bibitem{yamazawa_effect_2009}
Y.~Yamazawa,
\newblock Applied Physics Letters {\bf 95}, 191504 (2009).

\bibitem{rauf_effect_1998}
S.~Rauf and M.~J. Kushner,
\newblock Journal of applied physics {\bf 83}, 5087 (1998).

\bibitem{sobolewski_electrical_1999}
M.~A. Sobolewski and K.~L. Steffens,
\newblock Journal of Vacuum Science \& Technology A: Vacuum, Surfaces, and
  Films {\bf 17}, 3281 (1999).

\bibitem{kaiser_optimization_2013}
C.~Kaiser, D.~Maddex, M.~Pakala, and Q.~Leng,
\newblock Applied Physics Letters {\bf 103}, 232404 (2013).

\bibitem{yamazawa_control_2007}
Y.~Yamazawa, M.~Nakaya, M.~Iwata, and A.~Shimizu,
\newblock Japanese Journal of Applied Physics {\bf 46}, 7453 (2007).

\bibitem{yamazawa_electrode_2015}
Y.~Yamazawa,
\newblock Plasma Sources Science and Technology {\bf 24}, 034015 (2015).

\bibitem{verboncoeur_simultaneous_1993}
J.~P. Verboncoeur, M.~V. Alves, V.~Vahedi, and C.~K. Birdsall,
\newblock Journal of Computational Physics {\bf 104}, 321 (1993).

\bibitem{nagel_spice_1973}
L.~W. Nagel and D.~Pederson,
\newblock {SPICE} ({Simulation} {Program} with {Integrated} {Circuit}
  {Emphasis}),
\newblock Technical Report UCB/ERL M382, EECS Department, University of
  California, Berkeley, 1973.

\bibitem{vogt_ngspice_2017}
H.~Vogt, M.~Hendrix, and P.~Nenzi,
\newblock Ngspice {Users} {Manual} {Version} 27, 2017,
\newblock Available at: http://ngspice.sourceforge.net/docs/ngspice-manual.
  pdf.

\bibitem{mussenbrock_nonlinear_2006}
T.~Mussenbrock and R.~P. Brinkmann,
\newblock Applied Physics Letters {\bf 88}, 151503 (2006).

\bibitem{gudmundsson_notitle_2002}
J.~T. Gudmundsson,
\newblock Technical Report RH-21-2002, Science Institue, University of Iceland,
  2002.

\bibitem{godyak_situ_1990}
V.~A. Godyak and R.~B. Piejak,
\newblock Journal of Vacuum Science \& Technology A: Vacuum, Surfaces, and
  Films {\bf 8}, 3833 (1990).

\end{thebibliography}

\end{document}